\documentclass[12pt]{article}

\usepackage{eepic}
\usepackage{graphicx}
\usepackage{bbm}
\usepackage{amsmath}
\usepackage{alltt, amssymb}
\usepackage{amsfonts}
\usepackage{algorithmic}
\usepackage{algorithm}
\usepackage{pstricks}
\usepackage{graphicx} 
\usepackage{bbm}
\usepackage{amsmath,amsfonts,latexsym,amssymb}
\usepackage{algorithmic}
\usepackage{algorithm}

\def\A {\ensuremath{\mathbb{A}}}
\def\B {\ensuremath{\mathbb{B}}}

\def\D {\ensuremath{\mathbb{D}}}
\def\I {\ensuremath{\mathcal{I}}}
\def\J {\ensuremath{\mathcal{J}}}
\def\K {\ensuremath{\mathbf{k}}}
\def\L {\ensuremath{\mathbb{L}}}

\def\Q {\ensuremath{\mathbb{Q}}}

\def\KK {\ensuremath{\overline{\mathbf{k}}}}

\def\uu {\ensuremath{\mathbf{u}}}
\def\xx {\ensuremath{\mathbf{x}}}
\def\ss {\ensuremath{\mathbf{s}}}

\newtheorem{Lemma}{Lemma}
\newtheorem{Corollary}{Corollary}

\newtheorem{Example}{Example}
\newtheorem{Algorithm}{Algorithm}

\def\foorp{\hfill$\square$}

\def\myproof{\noindent{\small \sc Proof.}~}
\def\foorp{\hfill$\square$}

\newcommand{\lc}[1]{\mbox{{\rm lc}$(#1)$}}
\newcommand{\coeff}[1]{\mbox{{\rm coeff}$(#1)$}}

\newcommand{\dpol}[1]{\mbox{{\rm dpol}$(#1)$}}
\newcommand{\mat}[1]{\mbox{{\rm mat}$(#1)$}}
\newcommand{\fr}[1]{\mbox{{\rm fr}$(#1)$}}

\newcommand{\resultant}[1]{\mbox{{\sf res}$(#1)$}}

\newcommand{\init}[1]{\mbox{$\mathrm{init}(#1)$}}

\newcommand{\mvar}[1]{\mbox{$\mathrm{mvar}(#1)$}}
\newcommand{\prem}[1]{\mbox{$\mathrm{prem}(#1)$}}

\newcommand{\res}[1]{\mbox{{\rm res}$(#1)$}}
\newcommand{\sat}[1]{\mbox{$\mathrm{sat}(#1)$}}

\newcommand{\RegZero}[1]{\mbox{{\sf RegularizeDim0}$(#1)$}}
\newcommand{\RegInitZero}[1]{\mbox{{\sf RegularizeInitDim0}$(#1)$}}

\newcommand{\Regularize}[1]{\mbox{{\sf Regularize}$(#1)$}}

\newcommand{\Normalize}[1]{\mbox{{\sf Normalize}$(#1)$}}
\newcommand{\NormalForm}[1]{\mbox{{\sf NormalForm}$(#1)$}}

\newcommand{\RegularGcd}[1]{\mbox{{\sf RegularGcd}$(#1)$}}

\newcommand{\Maple}{{\sc Maple}}
\newcommand{\RegularChains}{{\tt RegularChains}}

\newcommand{\Modpn}{{\tt modpn}}

\newcommand{\RC}{{\tt RegularChains}}

\newcommand{\Magma}{\sc {Ma\-gma}}

\title{Computations Modulo Regular Chains}

\author{
Xin Li, Marc Moreno Maza and Wei Pan \\
{ORCCA, University of Western Ontario (UWO)} \\
London, Ontario, Canada \\
}

\begin{document}

\maketitle

\begin{abstract}
The computation of  triangular decompositions involves
two fundamental operations: 
polynomial GCDs modulo regular chains
and regularity test modulo saturated ideals.
We propose new algorithms for these core operations
based on modular methods and fast polynomial arithmetic.
We rely on new results connecting polynomial subresultants
and GCDs modulo regular chains.
We report on extensive experimentation, comparing our code to
pre-existing {\Maple} implementations, as well as more optimized {\Magma}
functions. In most cases, our new code outperforms the other
packages by several orders of magnitude.

\vspace*{0.2cm}
\end{abstract}


\section{Introduction}  %
\label{sect:Introduction}

A triangular decomposition of a set
$F \subset {\K}[x_1, \ldots, x_n]$
is a list of polynomial systems $T_1, \ldots, T_e$,
called {\em regular chains} (or regular systems) and
representing the zero set $V(F)$ of $F$.
Each regular chain $T_i$ may encode several irreducible components
of $V(F)$ provided that those share some properties (same dimension, 
same free variables, \ldots).

Triangular decomposition methods are based on a univariate and 
recursive vision of multivariate polynomials.
Most of their routines manipulate polynomial remainder
sequences (PRS).
Moreover, these methods  are usually ``factorization free'', which explains
why two different irreducible components
may be represented by the same regular chain.
An essential routine is then to check whether a hypersurface $f = 0$
contains one of the irreducible components encoded
by a regular chain $T$. This is achieved by testing whether
the polynomial $f$ is a zero-divisor modulo the so-called
{\em saturated ideal} of $T$.
The univariate vision on regular chains allows to perform this {\em regularity
test} by means of GCD computations. However, since the saturated
ideal of $T$ may not prime, the concept of a GCD used here is not standard.

The first formal definition of this type of GCDs was given
by Kalkbrener in \cite{Kalk93}.
But in fact, GCDs over non-integral domains 
were already used in several papers~\cite{DD87,Laz92,Gom92}
since the introduction
of the celebrated {\em D5 Principle}~\cite{D5}
by Della Dora, Dicrescenzo and Duval.
Indeed, this brilliant and simple observation allows
one to carry out over direct product of fields 
computations that are usually conducted over fields.
For instance, computing  univariate polynomial GCDs by means
of the Euclidean Algorithm.

To define a polynomial GCD
of two (or more) polynomials modulo a regular chain $T$,
Kalkbrener refers to the irreducible components that $T$ represents.
In order to improve the practical efficiency of those GCD computations by means
of subresultant techniques, Rioboo and the second author
proposed a more abstract definition in \cite{MMMRR}.
Their GCD algorithm is, however, limited to regular chains 
with zero-dimensional saturated ideals.

While Kalkbrener's definition cover the positive dimensional case,
his approach cannot support
triangular decomposition methods solving polynomial
systems incrementally,
that is, by solving one equation after another.
This is a serious limitation since incremental solving is a 
powerful way
to develop efficient sub-algorithms,
by means of geometrical consideration.
The first incremental triangular decomposition method was proposed
by Lazard in \cite{Laz91a}, without proof nor a GCD definition.
Another such method was established 
by the second author in~\cite{MMM99} together with a formal notion
of GCD adapted to the needs of incremental solving.
This concept, called {\em regular GCD}, is reviewed in
Section~\ref{sect:preliminary} 
in the context of regular chains.
A more abstract definition  follows.

Let {\B} be a commutative ring with unity.
Let $P, Q, G$ be non-zero univariate polynomials in ${\B}[y]$.
We say that $G$ is a {\em regular GCD} of $P, Q$ if the
following three conditions hold:
\begin{itemize}
\item[$(i)$] the leading coefficient of $G$ 
      is a regular element of {\B},
\item[$(ii)$]  $G$ lies in the ideal generated by 
      $P$ and $Q$ in ${\B}[y]$, and
\item[$(iii)$] if $G$ has positive degree w.r.t. $y$,
      then $G$ pseudo-divides both of $P$ and $Q$,
      that is, the pseudo-remainders \prem{P, G} and \prem{Q, G}
      are null.
\end{itemize}
In the context of regular chains, the ring {\B} is the
residue class ring of a polynomial ring ${\A} := {\K}[x_1, \ldots, x_n]$
(over a field {\K}) by the saturated ideal \sat{T} of a regular chain $T$.
Even if the leading coefficients of $P, Q$ are regular and
\sat{T} is radical, the polynomials $P, Q$ may not necessarily
admit a regular GCD (unless \sat{T} is prime). 
However, by splitting $T$ into several regular chains $T_1, \ldots, T_e$
(in a sense specified in Section~\ref{sect:preliminary})
one can compute a regular GCD of $P, Q$ 
over each of the ring ${\A} / \sat{T_i}$,
as shown in \cite{MMM99}.

In this paper, we propose a new algorithm for this task,
together with a theoretical study and implementation report,
providing dramatic improvements w.r.t. previous work~\cite{Kalk93,MMM99}.
Section~\ref{sec:RegularGcds} 
exhibits sufficient conditions for a subresultant 
polynomial of  $P,~Q \in {\A}[y]$ 
(regarded as univariate polynomials in $y$)
to be a regular GCD of  $P,~Q$  w.r.t. $T$.
Some of these properties could be known, but we could
not find a reference for them, in particular when \sat{T}
is not radical.
These results reduce the computation
of regular GCDs to that of subresultant chains, see  
Section~\ref{sec:RegularGcdAlgorithm} for details.

Since Euclidean-like algorithms tend to densify computations,
we consider 
an evaluation/interpolation scheme
based on FFT techniques for computing subresultant chains.
In addition, we observe that, while computing triangular
decomposition, whenever a regular GCD of 
$P$ and $Q$ w.r.t. $T$ is needed,
the resultant of $P$ and $Q$ w.r.t. $y$ is 
likely to be computed too.
This suggests to organize calculations
in a way that the subresultant chain of $P$ and $Q$ is computed only once.
Moreover, we wish to follow a successful principle  
introduced in~\cite{LMS07}: compute in 
${\K}[x_1, \ldots, x_n]$
instead of 
${\K}[x_1, \ldots, x_n] / \sat{T}$, as much as possible,
while controlling expression swell.
These three requirements targeting efficiency
are satisfied by the implementation techniques
of Section~\ref{sec:SpecilizationCube}.
The use of fast arithmetic for computing regular GCDs
was proposed in \cite{DMSX06} for 
regular chains with zero-dimensional radical saturated ideals.
However this method does not meet our other two requirements
and does not apply to arbitrary regular chains.
We state complexity results for the algorithms of this paper in
Sections~\ref{sec:SpecilizationCube} and 
\ref{sec:ResultantAndGCSolvingTwoEquations}.

Efficient implementation is the main objective
of our work.
We explain in Section~\ref{sec:ImplementationForRegulairze}
how we create
opportunities for using modular methods and
fast arithmetic in operations modulo regular chains,
such as regular GCD computation and regularity test.
The experimental results of 
Section~\ref{sec:experimentation}
illustrate the high efficiency of our algorithms.
We obtain speed-up factors of several orders
of magnitude w.r.t. the algorithms of  \cite{MMM99}
for regular GCD computations and regularity test.
Our code compares and often outperforms
packages with similar specifications in {\Maple}
and {\Magma}.


\section{Preliminaries}  %
\label{sect:preliminary}

Let $\K$ be a 
field and let $\K[\xx]=\K[x_1,\ldots,x_n]$ be the ring of polynomials
with coefficients in $\K$, with ordered variables $x_1 \prec \cdots \prec x_n$.
Let ${\KK}$ be the algebraic closure of $\K$.
If $\uu$ is a subset of $\xx$ then ${\K}(\uu)$ denotes the fraction field of
$\K[\uu]$.  
For $F \subset {\K}[\xx]$, we denote by $\langle F \rangle$ the
ideal it generates in ${\K}[\xx]$ and by $\sqrt{\langle F \rangle}$ the
radical of $\langle F \rangle$. For $H \in \K[\xx]$,
the {\em{saturated ideal}} of $\langle F \rangle$ w.r.t. $H$,
denoted by $\langle F \rangle:H^{\infty}$, is the ideal
$\{Q\in \K[\xx]\mid \exists m\in\mathbb{N}\text{ 
s.t. }H^mQ\in\langle F \rangle\}.$
A polynomial $P\in\K[\xx]$ is a {\em zero-divisor} modulo $\langle F
\rangle$ if there exists a polynomial $Q$ such that 
$P Q \in \langle F \rangle$, and neither $P$ nor $Q$ belongs to 
$\langle F \rangle$. The polynomial $P$ is {\em regular} modulo
$\langle F \rangle$ if it is neither zero, nor a zero-divisor modulo
$\langle F \rangle$. 
We denote by $V(F)$ the {\em zero set} (or algebraic 
variety) of $F$ in ${\KK}^n$.  For a subset $W \subset
{\KK}^n$, we denote by $\overline{W}$ its closure in the Zariski
topology.

\subsection{Regular chains and related notions}     %
\label{sect:preliminaryBasicDefinitions}

\smallskip\noindent{\small \bf Main variable and initial.}
If $P\in {\K}[\xx]$ is
a non-constant polynomial, the largest variable appearing in $P$ is
called the {\em main variable} of $P$ and is denoted by $\mvar{P}$.  
The leading coefficient of $P$ w.r.t. $\mvar{P}$ 
is its {\em initial}, written $\init{P}$
whereas $\lc{P,v}$ is the leading coefficient of $P$ w.r.t. $v \in \xx$.

\smallskip\noindent{\small \bf Triangular Set.}  
A subset $T$ of non-constant polynomials of ${\K}[\xx]$ 
is a {\em triangular set}
if the polynomials in $T$ have pairwise distinct main variables. 
Denote by $\mvar{T}$ the set of all  \mvar{P} for $P \in T$.
A variable $v\in\xx$ is {\em algebraic} w.r.t.
$T$ if $v \in \mvar{T}$; otherwise it is {\em free}.
For a variable $v \in \xx$ we denote by $T_{<v}$
(resp. $T_{>v}$) the subsets of $T$ consisting of the
polynomials with main variable less than (resp. greater than) $v$. 
If $v\in\mvar{T}$, we denote by $T_v$ the polynomial $ P \in T$
with main variable $v$. For $T$ not empty, 
$T_{\rm max}$ denotes the polynomial of $T$ with largest main variable.

\smallskip\noindent{\small \bf Quasi-component and saturated ideal.}
Given a triangular set $T$ in $\K[\xx]$, denote by $h_{T}$ the
product of the \init{P} for all $P \in T$.  
The {\em quasi-component} $W({T})$ of
$T$ is $V(T) \setminus  V(h_{T})$, that is, the set
of the points of $V({T})$ which do not cancel any of the initials of ${T}$.
We denote by \sat{T} the {\em saturated ideal} of $T$, defined
as follows: if $T$ is empty then \sat{T} is the trivial ideal
$\langle 0 \rangle$; otherwise it is the ideal 
$\langle T \rangle:h_{T}^{\infty}$. 

\smallskip\noindent{\bf \small Regular chain.}
A triangular set $T$ is a {\em regular chain} if either
$T$ is empty, or $T \setminus  \{ T_{max} \}$ is a regular
chain and the initial of ${T}_{max}$ is regular with
respect to $\sat{T \setminus  \{ T_{max} \}}$.  In this latter case,
$\sat{T}$ is a proper ideal of ${\K}[\xx]$.
From now on $T \subset {\K}[\xx]$ is a regular chain;
moreover we write $m=|T|$, $\ss= \mvar{T}$ and $\uu = \xx \setminus \ss$.
The ideal \sat{T} enjoys several properties.  First, its zero-set
equals $\overline{W(T)}$. Second, the ideal \sat{T} is unmixed
with dimension $n -m$. 
Moreover, any prime ideal ${\mathfrak{p}}$ associated to $\sat{T}$ 
satisfies ${\mathfrak{p}}\,{\cap}\,{\K}[\uu] = \langle 0 \rangle$.  
Third, if $n=m$, then $\sat{T}$ is simply $\langle T \rangle$.
Given $P \in {\K}[\xx]$ 
the {\em pseudo-remainder} (resp. {\em iterated resultant})
of $P$ w.r.t. $T$, denoted by \prem{P,T} (resp. \resultant{P,T})
is defined as follows. If $P \in {\K}$ or no variables of $P$ is 
algebraic w.r.t. $T$, then $\prem{P,T} = P$
(resp. $\resultant{P,T} = P$).
Otherwise, we set $\prem{P,T} = \prem{R,T_{<v}}$
(resp. $\resultant{P,T} = \resultant{R, T_{<v}}$)
where $v$ is the largest variable of $P$ which is algebraic w.r.t. $T$
and $R$ is the pseudo-remainder (resp. resultant) of $P$
and $T_v$ w.r.t. $v$.
We have:
$P$ is null (resp. regular) w.r.t. \sat{T} if and only if
$\prem{P,T} = 0$ (resp. $\resultant{P,T} \neq 0$).

\smallskip\noindent{\bf \small Regular GCD.}  Let $I$ be the
ideal generated by $\sqrt{\sat{T}}$ in ${\K}(\uu)[\ss]$. Then
${\L}(T) := {\K}(\uu)[\ss] /I$ is a direct product of fields. It
follows that every pair of univariate polynomials $P, Q \in {\L}(T)[y]$
possesses a GCD in the sense of \cite{MMMRR}.  The following 
GCD notion~\cite{MMM99} is convenient since it avoids
considering radical ideals.
Let $T \subset {\K}[x_1, \ldots, x_{n}]$ be a regular chain 
and let $P, Q \in {\K}[\xx,y]$ be non-constant polynomials 
both with main variable $y$. 
Assume that the initials of $P$ and $Q$ are regular modulo $\sat{T}$. 
A non-zero polynomial $G \in {\K}[\xx,y]$ is a {\em regular GCD} 
of $P, Q$ {\em w.r.t.} $T$ if these conditions hold:
\begin{itemize}
\item[$(i)$] $\lc{G,y}$ is regular with respect to $\sat{T}$;
\item[$(ii)$] there exist  $u, v \in {\K}[\xx,y]$ such that 
$g - u p - v t \in \sat{T}$;
\item[$(iii)$] if ${\deg}(G,y) > 0$ holds, then 
$\langle P, Q \rangle \subseteq \sat{T \cup G}$.
\end{itemize}

In this case, the polynomial $G$ has several properties. 
First, it is regular with respect to $\sat{T}$.  
Moreover, if $\sat{T}$ is radical and ${\deg}(G,y) > 0$ holds,
then the ideals $\langle P, Q \rangle$ and $\langle G \rangle$ of
${\L}({T})[y]$ are equal, so that $G$ is a GCD of $(P, Q)$
w.r.t. $T$ in the sense of~\cite{MMMRR}. 
The notion of a regular GCD can be used to compute intersections
of algebraic varieties.
As an example we will use Formula~(\ref{eq:RegularGCDFormula}) 
which follows from Theorem 32 in~\cite{MMM99}.
Assume that the regular chain $T$ is simply $\{ R \}$
where $R = \resultant{P,Q,y}$, for $ R \not\in {\K}$, 
and let $H$ be the product of the initials of $P$  and $Q$.
Then, we have:
\begin{eqnarray}
\label{eq:RegularGCDFormula}
V(P,Q) =  \overline{W(R,G)} \  \cup \  V(H,P,Q).
\end{eqnarray}

\smallskip\noindent{\bf \small Splitting.}  
Two polynomials $P, Q$ may not necessarily admit a regular GCD
w.r.t. a regular chain $T$, unless $\sat{T}$ is prime, see Example
$1$ in Section~\ref{sec:RegularGcds}. 
However, if $T$  ``splits'' into several regular chains,
then $P, Q$ may admit a regular GCD w.r.t. each of them. 
This requires a notation.
For non-empty regular chains 
$T, T_1, \ldots, T_e \subset {\K}[\xx]$
we write $T\longrightarrow (T_1, \ldots, T_e)$ 
whenever 
$\sqrt{\sat{T}} = \sqrt{\sat{T_1}}\cap \cdots\cap \sqrt{\sat{T_e}}$,
$\mvar{T} = \mvar{T_i}$ and $\sat{T}\subseteq\sat{T_i}$  hold 
for all $1 \leq i \leq e$.
If this holds, observe that any polynomial $H$ regular
w.r.t $\sat{T}$ is also regular w.r.t. $\sat{T_i}$ for all $1 \leq i \leq e$.

\subsection{Fundamental operations on regular chains}   %
\label{sect:preliminaryFundamentalOperations}

We list below the specifications of the fundamental operations on
regular chains used in this paper. The names and specifications
of these operations
are the same as in the {\RegularChains} library~\cite{LeMoXi05} in {\Maple}.

\smallskip 
\noindent{\small \bf Regularize.}
 For a regular chain $T \subset {\K}[\xx]$ and $P$ in ${\K}[\xx]$,
the operation $\Regularize{P,T}$ returns regular chains 
$T_1, \ldots, T_e$ of ${\K}[\xx]$ such that, for each $1 \leq i \leq e$,
$P$ is either zero or regular modulo
$\sat{T_i}$ and we have  $T {\longrightarrow} (T_1, \ldots, T_e)$.



\smallskip 
\noindent{\small \bf RegularGcd.}
Let $T$ be a regular chain and let $P, Q \in {\K}[\xx, y]$ be non-constant 
with $\mvar{P} = \mvar{Q} \not\in \mvar{T}$
and such that both $\init{P}$ and $\init{Q}$ are regular w.r.t.
$\sat{T}$. Then, the operation $\RegularGcd{P,Q,T}$ returns a
sequence 
$(G_1, T_1), \ldots, (G_e, T_e)$, called a {\em regular GCD sequence}, 
where $G_1, \ldots, G_e$ are polynomials 
and $T_1, \ldots, T_e$ are regular chains of
${\K}[\xx]$, such that $T {\longrightarrow} (T_1, \ldots, T_e)$
holds and $G_i$ is a regular GCD of $P, Q$ w.r.t. $T_i$ for all
$1 \leq i \leq e$.


\smallskip 
\noindent{\small \bf NormalForm.}
 Let $T$ be a zero-dimensional normalized regular chain, 
 that is, a regular chain whose saturated ideal
  is zero-dimensional and whose initials are all in the
 base field {\K}.
Observe that $T$ is a lexicographic Gr\"obner basis.
Then, for $P \in {\K}[\xx]$, the operation 
  \NormalForm{P,T} returns the {\em normal form} of $P$
     w.r.t. $T$ in the sense of Gr\"obner bases. 

\smallskip 
\noindent{\small \bf Normalize.}
Let $T$ be a regular chain
such that each variable occurring in $T$ belongs to \mvar{T}.
Let $P \in {\K}[\xx]$ be non-constant 
with initial $H$ regular w.r.t. $\langle T \rangle$.
Assume each variable of $H$ belongs to \mvar{T}.
Then $H$ is invertible modulo $\langle T \rangle$ and 
$\Normalize{P,T}$ returns $\NormalForm{H^{-1} P,T}$ where $H^{-1}$ is the
     inverse of $H$ modulo $\langle T \rangle$.

\subsection{Subresultants}    %
\label{ssec:Subresultants}

We follow the presentation of \cite{Duc97}, \cite{Yap1993}
and \cite{ElKahoui2003}.

\noindent{\small \bf Determinantal polynomial.}  Let {\B} be a
commutative ring with identity and let $m \leq n$ be positive 
integers.  Let $M$ be a $m \times n$ matrix with coefficients in {\B}.
Let $M_i$ be the square submatrix of $M$ consisting of the first $m-1$
columns of $M$ and the $i$-th column of $M$, for $i=m \cdots n$; let
$\det{M_i}$ be the determinant of $M_i$.  We denote by \dpol{M} the
element of ${\B}[y]$, called the {\em determinantal polynomial} of $M$,
given by
\begin{equation*}
  \det{M_m} y^{n-m} + \det{M_{m+1}} y^{n-m-1} + \cdots + \det{M_n}.
\end{equation*}
Note that if \dpol{M} is not zero then its degree is at most $n-m$.
Let $P_1, \ldots, P_m$ be polynomials of ${\B}[y]$ of degree less than
$n$.  We denote by $\mat{P_1, \ldots, P_m}$ the $m \times n$ matrix
whose $i$-th row contains the coefficients of $P_i$, sorting in order
of decreasing degree, and such that $P_i$ is treated as a polynomial
of degree $n-1$.  We denote by $\dpol{P_1, \ldots, P_m}$ the
determinantal polynomial of $\mat{P_1, \ldots, P_m}$.

\noindent{\small \bf Subresultant.}
Let $P, Q \in {\B}[y]$ be non-constant polynomials of respective
degrees $p,q$ with $q \leq p$. Let $d$ be an integer with $0 \leq d < q$.
Then the $d$-th {\em subresultant} of $P$ and $Q$, denoted
by $S_d(P,Q)$, is 
\begin{equation*}
\dpol{y^{q-d-1}P, y^{q-d-2}P, \ldots, P, y^{p-d-1}Q,\ldots, Q}.
\end{equation*}
This is a polynomial which belongs to the ideal generated 
by $P$ and $Q$ in ${\B}[y]$.
In particular, $S_0(P,Q)$ is \res{P,Q}, the resultant of $P$ and $Q$.
Observe that if $S_d(P,Q)$ is not zero then its degree is at most $d$.
When $S_d(P,Q)$ has degree $d$, it is said {\em non-defective} or 
{\em regular};
when $S_d(P,Q) \neq 0$ and ${\deg}(S_d(P,Q)) < d$, $S_d(P,Q)$ 
is said {\em defective}.
We denote by $s_d$ the coefficient of $S_d(P,Q)$ in $y^d$.
For convenience, we extend the definition to the $q$-th subresultant as
follows: 
\[
S_q(P,Q) = \left\{ 
\begin{array}{ll}
\gamma(Q)Q,       & \text{if } p>q \text{ or }\lc{Q}\in\B\text{ is regular} \\
\text{undefined}, &  otherwise
\end{array}
\right.
\]
where $\gamma(Q) = \lc{Q}^{p-q-1}$. Note that when $p$ equals $q$ and
$\lc{Q}$ is a regular element in $\B$, $S_q(P, Q)=\lc{Q}^{-1}Q$ is
in fact a polynomial over the total fraction ring of $\B$. 

We call {\em specialization property of subresultants} the following
statement. Let {\D} be another commutative ring with identity
and $\Psi$ a ring homomorphism from {\B} to {\D} such that
we have $\Psi(\lc{P}) \neq 0$ and $\Psi(\lc{Q}) \neq 0$.
Then we also have
\begin{equation*}
S_d(\Psi(P),\Psi(Q)) = \Psi(S_d(P,Q)).
\end{equation*}
             
{\small \bf Divisibility relations of subresultants.}
The subresultants $S_{q-1}(P,Q), S_{q-2}(P,Q)$, $\ldots$, $S_0(P,Q)$
satisfy relations which induce an Euclidean-like algorithm for computing them.
Following \cite{Duc97} we first assume that ${\B}$ is an integral domain.
In the above, we simply write $S_{d}$ instead of $S_{d}(P,Q)$, 
for $d=q-1, \ldots, 0$. We write $A \sim B$ for $A, B \in {\B}[y]$
whenever they are associated over $\fr{\B}$, the field of fractions of {\B}.
For $d =q-1,\ldots, 1$, we have:
\begin{itemize}
\item[$(r_{q-1})$] $S_{q-1} =  \prem{P,-Q}$, the pseudo-remainder of $P$ by $-Q$,
\item[$(r_{<q-1})$] if $S_{q-1} \neq 0$, with $e = {\deg}(S_{q-1})$, then
the following holds: $\prem{Q, - S_{q-1}} =  {\lc{Q}}^{(p-q)(q-e)+1} S_{e-1}$,
\item[$(r_e)$] if $S_{d-1} \neq 0$, with $e = {\deg}(S_{d-1}) < d-1$, 
               thus $S_{d-1}$ is defective,
               and we have
\begin{itemize}
\item[$(i)$] ${\deg}(S_{d}) = d$, thus $S_{d}$ is non-defective,
\item[$(ii)$] $S_{d-1} \sim S_e$ and 
              ${\lc{S_{d-1}}}^{d-e-1}S_{d-1} = {s_d}^{d-e-1} {S_e}$,
              thus $S_e$ is non-defective,
\item[$(iii)$] $S_{d-2} = S_{d-3} = \cdots = S_{e+1} = 0$,
\end{itemize}
\item[$(r_{e-1})$] if $S_d$ and $S_{d-1}$ are nonzero, with respective degrees 
                      $d$ and $e$, then we have
                  $\prem{S_d, - S_{d-1}} = {\lc{S_d}}^{d-e+1} S_{e-1}$,
\end{itemize}
We consider now the case where 
{\B} is an arbitrary commutative ring,
following Theorem 4.3 in~\cite{ElKahoui2003}.
If $S_d, S_{d-1}$ are non zero, with respective degrees 
$d$ and $e$ and if $s_d$ is regular in {\B}
then we have
${\lc{S_{d-1}}}^{d-e-1}S_{d-1} = {s_d}^{d-e-1} {S_e}$;
moreover, there exists $C_d \in {\B}[y]$ such that we have:
\begin{equation*}
(-1)^{d-1} \lc{S_{d-1}} s_e  S_d + C_d S_{d-1} =
{\lc{S_d}}^2  S_{e-1}.
\end{equation*}
In addition $S_{d-2} = S_{d-3} = \cdots = S_{e+1} = 0$ also holds.


\section{Regular GCDs}                %
\label{sec:RegularGcds}

Throughout this section, we assume $n \geq 1$ and we consider 
$P, Q \in {\K}[x_1,\ldots,x_{n+1}]$ non-constant polynomials with the
same main variable $y := x_{n+1}$ and such that
$p := {\deg}(P,y) \geq q := {\deg}(Q,y) $ holds.
We denote by $R$ the resultant of $P$ and $Q$ w.r.t. $y$.
Let $T \subset {\K}[x_1,\ldots,x_{n}]$ be a non-empty regular chain
such that $R \in \sat{T}$ and the  initials of $P, Q$ are regular
w.r.t. $\sat{T}$. 
We denote by ${\A}$ and ${\B}$ the rings ${\K}[x_1, \ldots, x_{n}]$
and ${\K}[x_1, \ldots, x_{n}] / \sat{T}$, respectively.
Let $\Psi$ be both the canonical ring homomorphism from ${\A}$ to ${\B}$
and the ring homomorphism it induces from ${\A}[y]$ to ${\B}[y]$.
For $0 \leq j \leq q$, we denote by $S_j$ the $j$-th
subresultant of $P, Q$ in ${\A}[y]$.

Let $d$ be an index in the range $1 \cdots q$ such that
$S_j \in \sat{T}$ for all $0 \leq j < d$.
Lemma~\ref{prop:RegularGcdModResultant5}
and Lemma~\ref{prop:RegularGcdModResultant3}
exhibit conditions under which $S_d$
is a regular GCD of $P$ and $Q$ w.r.t. $T$.
Lemma~\ref{prop:RegularGcdModResultant1}
and
Lemma~\ref{lemma::LSR}
investigate the properties of $S_d$
when $\mathrm{lc}(S_d, y)$ is regular modulo \sat{T}
and $\mathrm{lc}(S_d, y) \in \sat{T}$ respectively.

\begin{Lemma}
\label{prop:RegularGcdModResultant1}
If $\mathrm{lc}(S_d,y)$ is regular modulo $\sat{T}$, then
the polynomial $S_d$ is a non-defective subresultant of $P$ and $Q$ over $\A$.
Consequently,
$\Psi(S_d)$ is a non-defective subresultant of $\Psi(P)$ and
$\Psi(Q)$ in $\B[y]$.
\end{Lemma}
\myproof
When $d=q$ holds, we are done.
Assume $d<q$.
Suppose $S_d$ is defective, that is, $\deg(S_d, y) = e < d$.
According to item $(r_e)$ in the divisibility relations of subresultants,
there exists a non-defective subresultant $S_{d+1}$ such that
$$\lc{S_d,y}^{d-e} S_{d} = s_{d+1}^{d-e} S_e,$$
where $s_{d+1}$ is the leading coefficient of $S_{d+1}$ in $y$. 
By our assumptions, $S_e$ belongs to $\sat{T}$, thus 
$\lc{S_d,y}^{d-e} S_{d}\in\sat{T}$ holds. It follows from the fact
$\lc{S_d,y}$ is regular modulo $\sat{T}$
that $S_{d}$ is also in $\sat{T}$.
However the fact that $\lc{S_d,y} = \init{S_d}$ is regular modulo $\sat{T}$
also implies that $S_d$ is regular modulo $\sat{T}$. A contradiction.
\foorp

\begin{Lemma}
\label{lemma::LSR}
If $\mathrm{lc}(S_d, y)$ is contained in $\sat{T}$, 
then all the coefficients of $S_d$ regarded as a univariate polynomial in $y$
are nilpotent modulo $\sat{T}$.
\end{Lemma}
\myproof
If the leading coefficient $\lc{S_d, y}$ is in $\sat{T}$, 
then  $\lc{S_d, y}\in\mathfrak{p}$ holds for all the associated 
primes $\mathfrak{p}$ of $\sat{T}$.
By the Block Structure Theorem of subresultants 
(Theorem 7.9.1 of~\cite{Mis93}) over an integral domain 
$\K[x_1,\ldots,x_{n-1}]/\mathfrak{p}$, $S_d$ must belong to $\mathfrak{p}$. 
Hence we have $S_d\in\sqrt{\sat{T}}$.
Indeed, in a commutative ring,
the radical of an ideal equals  the intersection of all
its associated primes.
Thus $S_d$ is nilpotent modulo $\sat{T}$. 
It follows from Exercise 2 of~\cite{AM69} that
all the coefficients of $S_d$ in $y$ are also nilpotent modulo $\sat{T}$.
\foorp

Lemma~\ref{lemma::LSR} implies that,
whenever $\mathrm{lc}(S_d, y) \in \sat{T}$ holds,
the polynomial $S_d$ will vanish on all the components of $\sat{T}$ 
after splitting $T$ sufficiently. 
This is the key reason why Lemma~\ref{prop:RegularGcdModResultant1} 
can be applied for computing regular GCDs.
Indeed, up to splitting via the operation {\sf Regularize},
one can always assume that either
$\lc{S_d, y}$ is regular modulo $\sat{T}$
or $\lc{S_d, y}$ belongs to $\sat{T}$.
Hence, from Lemma~\ref{lemma::LSR} 
and up to splitting, one can assume that
either $\lc{S_d, y}$ is regular modulo $\sat{T}$
or $S_d$ belongs to $\sat{T}$.
Therefore, if $S_d \not\in \sat{T}$,
we consider the subresultant $S_d$ 
as a \emph{candidate regular GCD}
of $P$ and $Q$ modulo $\sat{T}$.

\begin{Example}
If $\mathrm{lc}(S_d,y)$ is not regular modulo $\sat{T}$ 
then $S_d$ may be defective.
Consider for instance the polynomials
$P = x_3^2 x_2^2-x_1^4$ and $Q = x_1^2 x_3^2 - x_2^4$
in $\mathbb{Q}[x_1, x_2, x_3]$. 
We have $\prem{P,-Q} =  (x_1^6 - x_2^6)$  and
$R = (x_1^6 - x_2^6)^2.$  Let $T = \{ R \}$.
The last subresultant of $P, Q$ modulo $\sat{T}$
is $\prem{P,-Q}$, which has degree 0 w.r.t $x_3$, although its index is $1$.
Note that $\prem{P,-Q}$ is nilpotent modulo $\sat{T}$.
\end{Example}

In what follows, we give sufficient conditions for 
the subresultant $S_d$ to be a regular GCD of $P$ and $Q$ w.r.t. $T$.
When $\sat{T}$ is a radical ideal, 
Lemma~\ref{prop:RegularGcdModResultant3}
states that the assumptions of Lemma~\ref{prop:RegularGcdModResultant1} 
are sufficient.
This lemma validates the search for a regular GCD of $P$ and $Q$
w.r.t. $T$ in a bottom-up style, 
from $S_0$ up to $S_{\ell}$ for some ${\ell}$.
Lemma~\ref{prop:RegularGcdModResultant5} covers the case 
where $\sat{T}$ is not radical
and states that $S_d$ is a regular GCD of 
$P$ and $Q$ modulo $T$,
provided that $S_d$ satisfies the conditions of 
Lemma~\ref{prop:RegularGcdModResultant1} and 
provided that, for all $d < k \leq q$, 
the coefficient $s_k$  of $y^k$ in $S_k$
is either null or regular modulo $\sat{T}$.

\begin{Lemma}
\label{prop:RegularGcdModResultant5}
We reuse the notations and assumptions of 
Lemma~\ref{prop:RegularGcdModResultant1}.
Then $S_d$ is a regular GCD of $P$ and $Q$ modulo $\sat{T}$, if for all 
$d < k \leq q$, the coefficient $s_k$ of $y^k$ in 
$S_k$ is either null or regular modulo $\sat{T}$.
\end{Lemma}
\myproof
There are three conditions to satisfy for $S_d$ to be a regular gcd of $P$ and $Q$ modulo $\sat{T}$:
\begin{itemize}
\item[$(1)$]  $\lc{S_d}$ is regular modulo $\sat{T}$;
\item[$(2)$]  there exists polynomials $u$ and $v$ such that $S_d - u P - v Q \in \sat{T}$; and
\item[$(3)$]  both $P$ and $Q$ are in $\I := \sat{T\cup\{S_d\}}$.
\end{itemize}
We will prove the lemma in three steps. We write $\Psi(r)$ as $\bar{r}$ for brevity
\footnote{We note that the degree of $\bar{S}_k$ may be less than the degree of $S_k$, since
its leading coefficient could be in $\sat{T}$. Hence, $\overline{\lc{S_k}}$ 
may differ from $\lc{\bar{S}_k}$. We carefully distinguish them when the leading
coefficient of a subresultant is not regular in $\B$.}.

{\small \sf Claim 1}: If $Q$ and $S_{q-1}$ are in $\sat{T}$,
then $S_d$ is a regular gcd of $P$ and $Q$ modulo $\sat{T}$.

Indeed, the properties of $S_d$ imply Conditions $(1)$ and $(2)$ 
and we only need to show that the Condition $(3)$ also holds. 
If $d=q$ holds, then $S_{q-1} \in \sat{T}$ and we are done.
Otherwise, $S_{q-1} = \prem{P, -Q}$ is not null modulo $\sat{T}$, 
because $\bar{S}_{q-1} = 0$ implies that all subresultants of $\bar{P}$ and $\bar{Q}$ 
with index less than $q$ vanish over $\B$.  
If both $S_{q} := Q$ and $S_{q-1} = \prem{P, -Q}$ are in $\I$, then
$P$ is also in $\I$, since $\lc{Q}$ is regular modulo $\sat{T}$ and hence is regular modulo $\I$.
This completes the proof of Claim 1.

In order to prove that $Q$ and $S_{q-1}$ are in $\sat{T}$, we
define the following set of indices
$$\J =  \{ j \mid  d<j<q, \coeff{S_j, y^j}\notin\sat{T}\}.$$
By assumption, $\coeff{S_j, y^j}$ is regular modulo $\sat{T}$ for each $j\in\J$.
Our arguments rely on the Block Structure Theorem (BST)
over an arbitrary ring~\cite{ElKahoui2003} and Ducos' subresultant algorithm~\cite{Duc97, MMM99}  
along with the specialization property of subresultants. 

{\small \sf Claim 2}: If $\J=\emptyset$, then $S_i\in\I$ holds 
for all $d<i\leq q$.

Indeed, the BST over $\B$ implies that there exists 
\emph{at most} one subresultant $S_j$ 
such that $d<j<q$ and $S_j\notin \sat{T}$. 
Therefore all but $S_{q-1}$ are in $\sat{T}$, and thus $\bar{S}_{q-1}$
is defective of degree $d$. More precisely, the BST over $\B$ implies
\begin{align}
\label{formula::blockRelation1}
\lc{\bar{S}_{q-1}}^{e} S_{q-1} & \equiv \lc{S_{q}}^{e} S_d \mod{\sat{T}}
\end{align}
for some integer $e\geq0$. 
According to Relation~(\ref{formula::blockRelation1}),
$\lc{\bar{S}_{q-1}}$ is regular in $\B$. Hence, we have $S_{q-1}\in \I$.
From the definition of $S_d$, 
we have $\prem{\bar{S_q}, -\bar{S}_{q-1}, y} \in \sat{T}$. 
This implies $S_q\in \I$.
This completes the proof of Claim 2.

Now we consider the case $\J \neq \emptyset$. Write $\J$ explicitly as 
$\J = \{ j_0, j_1, \ldots, j_{\ell-1} \},$ with $\ell=|\J|$ and we assume 
$j_0<j_1<\cdots<j_{\ell-1}$. For convenience, we write $j_{\ell} := q$.
For each integer $k$ satisfying $0\leq k\leq \ell$
we denote by ${\cal P}_k$ the following property:
\begin{equation*}
S_i \in \I,  \ \ {\rm for} \ \ {\rm all} \ \ d<i\leq j_{k}.
\end{equation*}

{\small \sf Claim 3}: 
The property ${\cal P}_k$ holds for all $0\leq k\leq \ell$.

We proceed by induction on $0\leq k\leq {\ell}$. 
The base case is $k=0$. We need to show $S_i \in \I$ for all $d < i \leq j_0$.
By the definition of $j_0$, $\bar{S}_{j_0}$ is a non-defective subresultant 
of $\bar{P}$ and $\bar{Q}$, and $\coeff{S_i, y^i}$ is in $\sat{T}$ for all $d<i<j_0$.
By the BST over $\B$, there is \emph{at most} one $d<i<j_0$ such that $S_i\notin\sat{T}$.
If no such a subresultant exists, then we know that 
$\prem{\bar{S}_{j_0}, -\bar{S}_d}$ is in $\sat{T}$.
Consequently, $S_{j_0}\in\I$ holds, 
which implies $S_j\in\I$ for all $d<i\leq j_0$.
On the other hand, if $S_{i_0}$ is not in $\sat{T}$ for some $d<i_0<j_0$, then 
$\bar{S}_{i_0}$ is similar to $\bar{S}_d$ over $\B$. 
To be more precise, we have
\begin{align}
    \lc{ \bar{S}_{i_0}}^{e} \bar{S}_{i_0} \equiv \lc{\bar{S}_{j_0}}^{e}\bar{S}_d \mod{\sat{T}}
\end{align}
for some integer $e\geq0$. 
With the same reasoning as in the case $\J=\emptyset$, we know that
$\lc{\bar{S}_{i_0}}$ is regular modulo $\sat{T}$ and we deduce that 
$S_{i_0}\in \I$ holds. 
Also, we have $\prem{\bar{S}_{j_0}, -\bar{S}_{i_0}} \in \sat{T}$, 
by definition of $S_d$. 
This implies $S_{j_0}\in\I$ from the fact
that $\lc{\bar{S}_{i_0}}$ is regular modulo $\sat{T}$ 
(and thus regular modulo $\I$).
Hence,  we have $S_{i} \in \I$ for all $d<i\leq j_0$, as desired.
Therefore the property ${\cal P}_{k}$ holds for $k=0$.

Now we assume that the property ${\cal P}_{k-1}$ holds for 
some $1 \leq k\leq \ell$.
We prove that ${\cal P}_{k}$ also holds.
According to the BST over $\B$, we know that there exists 
\emph{at most} one subresultant between $\bar{S}_{j_{k-1}}$ and $\bar{S}_{j_k}$, 
both of which are non-defective subresultants of $\bar{P}$ and $\bar{Q}$.
If $S_{i}\in \sat{T}$ holds for all $ j_{k-1} < i <  j_k$, then we have
$$\prem{\bar{S}_{j_k}, -\bar{S}_{j_{k-1}}} \equiv \lc{\bar{S}_{j_k}}^{e}\bar{S}_u \mod{\sat{T}}$$ 
for some $d \leq u < j_{k-1}$ and some integer $e\geq0$.
Thus, we have $\prem{\bar{S}_{j_k}, -\bar{S}_{j_{k-1}}} \in \I$ 
by our induction hypothesis,
and consequently, $S_{j_k} \in \I$ holds. 
On the other hand, if all subresultants $S_i$ (for $j_{k-1}<i<j_k$)
but $S_{i_k}$ (for some index $i_k$ such that $j_{k-1}<i_k<j_k$)
are in $\sat{T}$, then $\bar{S}_{i_k}$ is similar to $\bar{S}_{j_{k-1}}$ 
over $\B$, 
that is, we have
\begin{align}
    \label{formula::blockRelation3}
    \lc{ \bar{S}_{i_k}}^{e} \bar{S}_{i_k} \equiv \lc{\bar{S}_{j_{k}}}^{e}\bar{S}_{j_{k-1}} \mod{\sat{T}}
\end{align}
for some integer $e\geq0$. By Relation~(\ref{formula::blockRelation3}), 
$\lc{\bar{S}_{i_k}}$ is regular modulo $\sat{T}$, 
and thus is regular modulo $\I$. 
Using Relation~(\ref{formula::blockRelation3}) again, 
we have $S_{i_k}\in\I$, 
since $S_{j_{k-1}}$ is in $\I$. 
Also, we have 
$$\prem{\bar{S}_{j_k}, -\bar{S}_{i_k}} \equiv \lc{\bar{S}_{j_k}}^{e} \bar{S}_{u} \mod{\sat{T}}$$
for some $d\leq u<j_{k-1}$ 
and some integer $e\geq0$. 
By the induction hypothesis, 
we deduce $S_u\in\I$, which implies $S_{j_k} \in \I$ 
together with the fact that $\lc{\bar{S}_{i_k}}$ is regular modulo $\I$.
This shows that $S_i\in\I$ holds for all $d<i\leq j_{k}$. 
Therefore, property ${\cal P}_{k}$ holds.

Finally, we apply Claim 3 with $k=\ell$, leading to
$S_i\in\I$ for all $d< i\leq j_{\ell} = q$, 
which completes the proof of our lemma.
\foorp

\begin{figure}[htbp]
\centering
\psset{unit=0.7cm}
\begin{pspicture}(0,0)(20,8)
\psline[linewidth=0.8mm,linecolor=gray](7,0)(8,0)
\put(8.2,-0.2){\scriptsize $S_1$}
\psline[linewidth=0.8mm,linecolor=gray](6,1)(8,1)
\put(8.2,0.9){\scriptsize $S_{2}$}
\psline[linewidth=0.8mm,linecolor=gray](4,3)(8,3)
\put(8.2,2.9){\scriptsize $S_{4}$}
\psline[linewidth=0.8mm,linecolor=gray](4,4)(8,4)
\put(8.2,3.9){\scriptsize $S_{5}$}
\psline[linewidth=0.8mm,linecolor=gray](2,5)(8,5)
\put(8.2,4.9){\scriptsize $S_{6}$}
\psline[linewidth=0.8mm,linecolor=gray](1,6)(8,6)
\put(8.2,5.9){\scriptsize $Q=S_{7}$}
\psline[linewidth=0.8mm,linecolor=gray](0,7)(8,7)
\put(8.2,6.9){\scriptsize $P$}
\psline[linestyle=dotted,linecolor=gray](7,0)(1,6)
\psline[linewidth=0.8mm,linecolor=gray](17,0)(18,0)
\put(18.2,-0.2){\scriptsize $\bar{S}_1$}
\psline[linewidth=0.8mm,linecolor=gray](17,1)(18,1)
\put(18.2,0.9){\scriptsize $\bar{S}_{2}$}
\psline[linewidth=0.8mm,linecolor=gray](14,3)(18,3)
\put(18.2,2.9){\scriptsize $\bar{S}_{4}$}
\psline[linewidth=0.8mm,linecolor=gray](14,5)(18,5)
\put(18.2,4.9){\scriptsize $\bar{S}_{6}$}
\psline[linewidth=0.8mm,linecolor=gray](11,6)(18,6)
\put(18.2,5.9){\scriptsize $\bar{Q}=\bar{S}_{7}$}
\psline[linewidth=0.8mm,linecolor=gray](10,7)(18,7)
\put(18.2,6.9){\scriptsize $\bar{P}$}
\psline[linestyle=dotted,linecolor=gray](17,0)(11,6)
\end{pspicture}
\caption{ 
    \textrm{A possible configuration of the subresultant chain of $P$ and $Q$. In the left, $P$ and $Q$ have
    five nonzero subresultants over $\K[\xx]$, four of which are non-defective and one of which is defective.
    Let $T$ be a regular chain in $\K[\xx]$ such that $\lc{P}$ and $\lc{Q}$ are regular modulo $\sat{T}$. 
    Further, we assume that $\lc{S_1}$ and $\lc{S_4}$ are regular modulo $\sat{T}$, however, $\lc{S_6}$ is in $\sat{T}$.
    The right hand side is a possible configuration of the subresultant chain of $\bar{P}$ and $\bar{Q}$.
    In the proof of Claim 3, the set $\J$ is $\{j_0 = 4\}$ and $j_1=7$, whereas $i_0=2$ and $i_1=6$ are the indices of
    defective subresultants over $\K[\xx]/\sat{T}$. In this case, $S_1$ is a regular gcd of $P$ and $Q$ modulo $\sat{T}$.}
}
\label{fig:BST}
\end{figure}
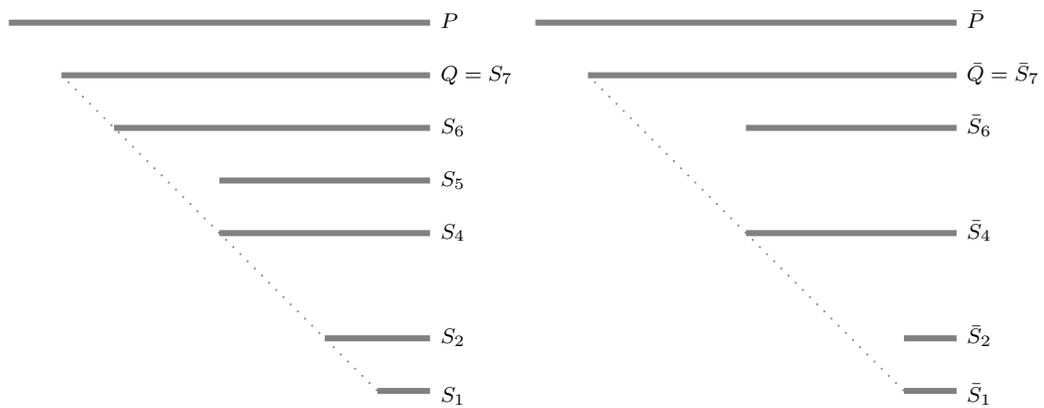

The consequence of the above corollary is that we ensure that $S_d$ 
is a regular gcd after checking that 
the leading coefficients of all non-defective subresultants above $S_d$,
are either null or regular modulo $\sat{T}$.
Therefore, one may be able to conclude that $S_d$ is a regular GCD
simply after checking the coefficients ``along the diagonal''
of the pictorial representation of the subresultants of $P$ and $Q$,
see Figure~\ref{fig:BST}.

\begin{Lemma}
\label{prop:RegularGcdModResultant3}
With the assumptions of 
Lemma~\ref{prop:RegularGcdModResultant1},
assume \sat{T} radical.
Then, $S_d$ is a regular GCD of $P, Q$ w.r.t. $T$.
\end{Lemma}
\myproof
As for Lemma~\ref{prop:RegularGcdModResultant5},
it suffices to check that  $P$ and $Q$ belong to 
$\sat{ T \cup \{ S_d \}}$.
Let ${\mathfrak{p}}$ be any prime ideal associated with \sat{T}.
Define ${\D} = {\K}[x_1, \ldots, y] / {\mathfrak{p}}$
and let {\L} be the fraction field of the integral domain {\D}.
Clearly $S_d$ is the last subresultant of $P,~Q$ in ${\D}[y]$
and thus in ${\L}[y]$.
Hence $S_d$ is a GCD of $P,~Q$ in ${\L}[y]$.
Thus $S_d$ divides $P,~Q$ in ${\L}[y]$
and pseudo-divides $P,~Q$ in ${\D}[y]$.
Therefore $\prem{P,S_d}$ and $\prem{Q,S_d}$
belong to ${\mathfrak{p}}$.
Finally $\prem{P,S_d}$ and $\prem{Q,S_d}$ belong to \sat{T}.
Indeed, \sat{T} being radical, it is the intersection
of its associated primes.
\foorp

\section{A regular GCD algorithm}     %
\label{sec:RegularGcdAlgorithm}
\label{sec:RGSZR}

Following the notations and assumptions of 
Section~\ref{sec:RegularGcds}
we propose an algorithm for 
computing a regular GCD sequence of $P,Q$ w.r.t. $T$.
as specified in Section~\ref{sect:preliminaryFundamentalOperations}.
Then, we show how to relax the assumption $R \in \sat{T}$.

There are three main ideas behind this  algorithm.
First, the subresultants of $P,Q$ in ${\A}[y]$ are assumed to be known.
We explain in Section~\ref{sec:ImplementationAndComplexity}
how we compute them in our implementation.
Secondly, we rely on the  {\small \bf Regularize}
operation specified 
in Section~\ref{sect:preliminaryFundamentalOperations}.
Lastly, we inspect the subresultant chain of $P,Q$ in ${\A}[y]$
in a bottom-up manner.
Therefore, we view $S_1, S_2, \ldots$ as successive candidates
and apply either Lemma~\ref{prop:RegularGcdModResultant3},
(if $\sat{T}$ is known to be  radical) or 
Lemma~\ref{prop:RegularGcdModResultant5}.

\smallskip\noindent{\small \bf Case where $R \in \sat{T}$.}   %
By virtue of Lemma~\ref{prop:RegularGcdModResultant1}
and
Lemma~\ref{lemma::LSR} there exist regular chains
 $T_1, \ldots, T_e \subset {\K}[\xx]$
such that  $T\longrightarrow (T_1, \ldots, T_e)$ holds
and 
for each $1 \leq i \leq e$
there exists an index $1 \leq d_i \leq q$ such that
the leading coefficient 
$\lc{S_{d_i},y}$ of the subresultant $S_{d_i}$
is regular modulo $\sat{T}$ and 
$S_j \in \sat{T_i}$  for all $0 \leq j < d_i$.
Such regular chains can be computed using the operation
{\sf Regularize}.
If each $\sat{T_i}$ is radical then it follows
from Lemma~\ref{prop:RegularGcdModResultant3}
that $(S_{d_1}, T_1), \ldots, (S_{d_e}, T_e)$ 
is a regular GCD sequence of $P, Q$ w.r.t. $T$.
In practice, when \sat{T} is radical then 
so are all $\sat{T_i}$, see \cite{BLM06}.
If some $\sat{T_i}$ is not known to be radical,
then one can compute regular chains
 $T_{i,1}, \ldots, T_{i,e_i} \subset {\K}[\xx]$
such that  $T_i \longrightarrow (T_{i,1}, \ldots, T_{i,e_i})$ holds
and 
for each $1 \leq {\ell}_i \leq e_i$
there exists an index $1 \leq d_{{\ell}_i} \leq q$ such that
Lemma~\ref{prop:RegularGcdModResultant5} applies
and shows that the subresultant $S_{d_{{\ell}_i}}$ is
regular GCD of $P, Q$ w.r.t. $T_{i,{\ell}_i}$.
Such computation relies again on 
{\sf Regularize}.

\smallskip\noindent{\small \bf Case where $R \not\in \sat{T}$.} %
\label{sec:RegularGcdAlgorithmGeneral}
We explain how to relax the assumption $R \in \sat{T}$
and thus obtain a general algorithm for the operation $\mathsf{RegularGcd}$.
The principle is straightforward. 
Let $R=\resultant{P,Q, y}$. 
We call $\Regularize{R, T}$ obtaining regular chains
$T_1, \ldots, T_e$ such that $T \longrightarrow (T_1, \ldots, T_e)$.
For each $1 \leq i \leq e$ we compute a regular GCD sequence of
$P$ and $Q$ w.r.t. $T_i$ as follows:
If $R \in \sat{T_i}$ holds then we proceed as described above;
otherwise $R \not\in \sat{T_i}$ holds 
and the resultant $R$ is actually a regular
GCD of $P$ and $Q$ w.r.t. $T_i$ by definition.
Observe that when $R \in \sat{T_i}$ holds
the subresultant chain of $P$ and $Q$ in $y$ is used
to compute their regular GCD w.r.t. $T_i$.
This is one of the motivations for the implementation
techniques described in Section~\ref{sec:ImplementationAndComplexity}.


\section{Implementation and Complexity}  %
\label{sec:ImplementationAndComplexity}

In this section we address implementation techniques
and complexity issues.
We follow the notations introduced in Section~\ref{sec:RegularGcds}.
However we do not assume that $R  = \resultant{P,Q,y}$
belongs to the saturated ideal of the regular chain $T$.

In Section~\ref{sec:SpecilizationCube} we describe
our encoding of the subresultant chain of $P, Q$
in ${\K}[\xx][y]$.
This representation is used in our
implementation and complexity results.
For simplicity our analysis is restricted to the
case where {\K} is a finite field 
whose ``characteristic is large enough''.  
The case where {\K} is the field {\Q} of rational numbers
could be handled in a similar fashion, with the necessary adjustments.

One motivation for the design of the techniques presented in this paper is 
the solving of systems of two equations, say $P = Q = 0$.
Indeed, this can be seen as a fundamental operation 
in incremental methods for solving systems of polynomial equations,
such as the one of \cite{MMM99}.
We make two simple observations.
Formula~\ref{eq:RegularGCDFormula} p.~\pageref{eq:RegularGCDFormula}
shows that solving this system reduces ``essentially''
to computing $R$ and a regular GCD sequence of $P,~Q$ modulo
$\{ R \}$, when $R$ is not constant.
This is particularly true when $n = 2$ since in this case 
the variety $V(H,P,Q)$ is likely to be empty 
for ``generic'' polynomials $P, Q$.
The second observation is that, under the same genericity 
assumptions, a regular GCD $G$ 
of $P, Q$ w.r.t. $\{ R \}$ is likely to exist
and have degree one w.r.t. $y$.
Therefore, once the subresultant chain of $P, Q$ w.r.t. $y$
is calculated, one can obtain $G$ ``essentially'' at no additional cost.
Section~\ref{sec:ResultantAndGCSolvingTwoEquations}
extends these observations with complexity results.

In Section~\ref{sec:ImplementationForRegulairze}
an algorithm for  {\sf Regularize} and
its implementation are discussed.
We show how to create opportunities
for using fast polynomial arithmetic
and modular techniques, thus bringing significant
improvements w.r.t. other algorithms for the same operation,
as shown in Section~\ref{sec:experimentation}.

\subsection{Subresultant chain encoding} %
\label{sec:SpecilizationCube}

Following~\cite{Coll71}, we evaluate $(x_1, \ldots, x_{n})$ at sufficiently
may points such that the subresultants of $P$ and $Q$ (regarded
as univariate polynomials in $y = x_{n+1}$) can be computed
by interpolation.
To be more precise, we need some notations.
Let $d_i$ be the maximum of the degrees of $P$ and $Q$ 
in $x_i$, for all $i = 1, \ldots, n+1$. 
Observe that $b_i := 2 d_i d_n$ is an upper bound for
the degree of $R$ (or any subresultant of $P$ and $Q$) in $x_i$, for
all $i$. Let $B$ be the product $(b_1 +1)
\cdots (b_{n} + 1)$.

We proceed by evaluation / interpolation; our sample points are chosen
on an $n$-dimensional rectangular grid. 
We call ``Scube'' the values of the
subresultant chain of $P,Q$ on this grid, 
which is precisely 
how the subresultants of $P,Q$ are encoded in our implementation.
Of course, the validity of
this approach requires that our evaluation points cancel no initials 
of $P$ or $Q$.
Even though finding such points deterministically
is a difficult problem, this created no issue in our implementation.
Whenever possible (typically, over suitable finite fields), we choose
roots of unity as sample points, so that we can use FFT (or van der
Hoeven's Truncated Fourier Transform~\cite{Hoeven04}); otherwise, the
standard fast evaluation / interpolation algorithms are used. 
We have $O(d_{n+1})$ evaluations and $O(d_{n+1}^2)$ interpolations to perform. 
Since our sample points lie on a grid, the total cost becomes
$$O\left ( B d^2_{n+1} \sum_{i=1}^{n} \log(b_i) \right ) \quad
\text{or}\quad O\left ( B d^2_{n+1}
\sum_{i=1}^{n}\frac{\mathsf{M}(b_i)\log(b_i)}{b_i} \right ),$$
depending on the choice of the sample points (see e.g.~\cite{Pan94}
for similar estimates). 
Here, as usual, $\mathsf{M}(b)$ stands for the
cost of multiplying polynomials of degree less than $b$,
see~\cite[Chap.~8]{GaGe99}. Using the estimate $\mathsf{M}(b) \in O(b
\log(b)\log\log(b))$ from~\cite{CantorKaltofen1991}, this respectively
gives the bounds
$$O (  d^2_{n+1} B \log(B) ) \quad\text{and}\quad
  O (  d^2_{n+1} B \log^2(B)\log\log(B) ).$$
These estimates are far from optimal. 
A first important improvement (present in
our code) consists in interpolating in the  first place only the {\it leading
coefficients} of the subresultants, and recover all
other coefficients when needed. This is sufficient for the 
algorithms of Section~\ref{sec:RegularGcds}.
For instance, in the FFT case, the cost is reduced to 
$$O ( d^2_{n+1} B + d_{n+1} B \log(B) ).$$
Another desirable improvement would of course consist in using fast
arithmetic based on {\em Half-GCD} techniques~\cite{GaGe99}, with the
goal of reducing the total cost to $O\tilde{~}( d_{n+1} B )$, which is the
best known bound for computing the resultant, or a given
subresultant. However, as of now, we do not have such a result, due to
the possible splittings.

\subsection{Solving two equations}  %
\label{sec:ResultantAndGCSolvingTwoEquations}

Our goal now is to estimate the cost of computing the polynomials $R$
and $G$ in the context of  
Formula~\ref{eq:RegularGCDFormula} p.~\pageref{eq:RegularGCDFormula}.
We propose an approach where the computation of $G$ essentially comes
come free, once $R$ has been computed.  This is a substantial
improvement compared to traditional methods, such
as~\cite{Kalk93,MMM99}, which compute $G$ without recycling 
the intermediate calculations of $R$.
With the assumptions and notations of
Section~\ref{sec:SpecilizationCube}, we saw that the 
resultant $R$ can be computed 
in at most $O(d_{n+1} B {\log}(B) + d^2_{n+1} B)$ operations in {\K}.
In many cases (typically, with random systems), $G$ has degree one in
$y =x_{n+1}$. Then, the GCD $G$ can be computed within the same bound as
the resultant. Besides, in this case, one can use the Half-GCD
approach instead of computing all subresultants of $P$ and $Q$. This
leads to the following result in the bivariate case; we omit its proof here.
\begin{Corollary}
\label{prop:OurAlgorithm2}
With $n=2$, if $V(H, P, Q)$ is empty and ${\deg}(G,y) = 1$, then
solving the input system $P = Q = 0$ can be done in $O^{\sim}(d^2_2
d_1)$ operations in {\K}.
\end{Corollary}


\subsection{Implementation of Regularize}    %
\label{sec:ImplementationForRegulairze}

The operation {\sf Regularize} specified in 
Section~\ref{sect:preliminaryBasicDefinitions}
is a core routine in methods computing triangular decompositions. 
It has been used in the algorithms
presented in Section~\ref{sec:RegularGcdAlgorithm}.
Algorithms for this operation appear in \cite{Kalk93,MMM99}.

The purpose of this section is to show how to realize
efficiently this operation.
For simplicity, we restrict ourselves to 
regular chains with zero-dimensional saturated ideals,
in which case the {\sf separate} operation of \cite{Kalk93}
and the {\sf regularize} operation \cite{MMM99} are similar.
For such a regular chain $T$ in ${\K}[\xx]$ 
and a polynomial $P \in {\K}[\xx]$ 
we denote by $\RegZero{P,T}$ the function call $\Regularize{P, T}$.
In broad terms, it ``separates'' the points of
$V(T)$ that cancel $P$ from those which do not.  
The output is a set of regular chains $\{T_1, \ldots, T_e\}$ such that
the points of $V(T)$ which cancel $p$ are given by 
the $T_i$'s modulo which $p$ is null.

Algorithm~\ref{Algo:IsInvertible} differs from those 
with similar specification in~\cite{Kalk93,MMM99}
by the fact it creates opportunities for using
modular methods and fast polynomial arithmetic.
Our first trick is based on the following result 
(Theorem 1 in~\cite{CLGMP07}):
the polynomial $p$ is invertible modulo $T$ if and only
if the iterated resultant of $P$ with respect to $T$ is non-zero.
The correctness of Algorithm~\ref{Algo:IsInvertible} follows from 
this result, the specification of the operation {\sf RegularGcd}
and an inductive process. Similar proofs appear in~\cite{Kalk93,MMM99}.
A proof and complexity analysis of Algorithm~\ref{Algo:IsInvertible}
will be reported in another article.

The main novelty of Algorithm~\ref{Algo:IsInvertible} is to
employ the fast evaluation/interpolation strategy
described in Section~\ref{sec:SpecilizationCube}.
In our implementation of Algorithm~\ref{Algo:IsInvertible}, 
at Step $(6)$, we compute the ``Scube'' representing
the subresultant chain of $q$ and $C_v$.
This allows us to compute the resultant $r$ and then to compute 
the regular GCDs $(g, E)$ at Step $(12)$ 
from the same ``Scube''. In this way, 
intermediate computations are recycled.
Moreover, fast polynomial arithmetic is involved 
through the manipulation of the ``Scube''.

\begin{Algorithm} 
\label{Algo:IsInvertible}
\noindent \begin{description}
\item[{\bf Input:}] $T$ a normalized zero-dimensional regular chain and $P$
                    a polynomial, both in ${\K}[x_1, \ldots, x_n]$.
\item[{\bf Output:}] See specification in Section~\ref{sect:preliminaryFundamentalOperations}.
\end{description}
\begin{tabbing}
\quad \= \quad \= \quad \= \quad \= \quad \= \quad \= \quad \= \quad \= \quad \=  \quad \= \quad \= \quad \= \quad \= \quad \kill
$\RegZero{P,T}$ == \\
$(1)$ \> \> $Results := \emptyset$;\\
$(2)$ \> \> {\bf for} $(q, C)\in\RegInitZero{P,T}$ {\bf do} \\
$(3)$ \> \> \> \> {\bf if} $q \in {\K}$ {\bf then} \\ 
$(4)$ \> \> \> \> \> \> $Results$ := $\{C\} \cup Results$ \\
$(5)$ \> \> \> \> {\bf else} $v$ := $\mvar{q}$ \\
$(6)$ \> \> \> \> \> \> ~$r$ := $\resultant{q,C_v,v}$ \\
$(7)$ \> \> \> \> \> \> ~{\bf for} $D\in\RegZero{r, C_{<v}}$ {\bf do} \\
$(8)$ \> \> \> \> \> \> \> \> $s := \NormalForm{r, D}$ \\
$(9)$ \> \> \> \> \> \> \> \> {\bf if} $s\neq0$ {\bf then}\\ 
$(10)$ \> \> \> \> \> \> \> \> \> \> $U := \{ D \cup \{C_v\} \cup C_{>v}\}$ \\
$(11)$ \> \> \> \> \> \> \> \> \> \> $Results := \{ U \} \cup Results$ \\
$(12)$ \> \> \> \> \> \> \> \> {\bf else for} $(g, E)\in\RegularGcd{q, C_v, D}$ {\bf do}\\
$(13)$ \> \> \> \> \> \> \> \> \> \> $g := \NormalForm{g, E}$ \\
$(14)$ \> \> \> \> \> \> \> \> \> \> $U := \{ E \cup \{g\} \cup D_{>v}\}$ \\
$(15)$ \> \> \> \> \> \> \> \> \> \> $Results := \{U\} \cup Results$\\
$(16)$ \> \> \> \> \> \> \> \> \> \> $c := \NormalForm{{\rm quo}(C_{v}, g), E}$ \\
$(17)$ \> \> \> \> \> \> \> \> \> \> {\bf if} $\deg(c,v) > 0$ {\bf then} \\
$(18)$ \> \> \> \> \> \> \> \> \> \> \> \> $Results := \RegZero{q, E \cup c \cup C{_>v}}\cup\, Results$\\ 
$(19)$ \> \> {\bf return} $Results$
\end{tabbing}
\end{Algorithm}
In Algorithm~\ref{Algo:IsInvertible}, 
a routine {\sf RegularizeInitialDim0} is called, 
whose specification is
given below. See~\cite{MMM99} for an algorithm.


\begin{description}
\item[{\bf Input:}] $T$ a normalized zero-dimensional regular chain and $p$
                    a polynomial, both in ${\K}[x_1, \ldots, x_n]$.
\item[{\bf Output:}] A set of pairs $\{(p_i, T_i) \mid i=1\cdots e\}$,
in which $p_i$ is a polynomial and $T_i$ is a regular chain, such that
either $p_i$ is a constant or its initial is regular modulo $\sat{T_i}$, and $p \equiv p_i \bmod \sat{T_i}$ holds; 
moreover we have $T\longrightarrow (T_1, \ldots, T_e)$.
\end{description}

\section{Experimentation}
\label{sec:experimentation}

We have implemented in C language all the algorithms presented
in the previous sections. The corresponding functions rely on 
the asymptotically fast  arithmetic operations from 
our {\Modpn} library~\cite{LMRS08}.
For this new code, we have also realized a {\Maple} interface,
called {\sf FastArithmeticTools},
which is a new module of the {\RC} library~\cite{LeMoXi05}.

In this section, we compare the performance of our 
{\tt Fast\-ArithmeticTools} commands with
{\Maple}'s and {\Magma}'s existing counterparts.
For {\Maple}, we use its latest release, namely 
version 13; 
For {\Magma} we use Version {\tt V2.15-4}, which is the latest one
at the time of writing this paper.
However, for this release, the {\Magma} commands 
{\tt Triangular\-Decomposition} and {\tt Saturation}
appear to be some time much slower than in Version {\tt V2.14-8}.
When this happens, we provide timings for both versions.

We have three test cases
dealing respectively with the solving of bivariate systems,
the solving of systems of two equations and 
the regularity testing of a polynomial w.r.t. a zerodimensional
regular chain.
In our experimentation all polynomial coefficients
are in a prime field whose characteristic is a 30bit
prime number. For each of our figure or table
the ``degree'' is the total degree of any polynomial
in the input system.
All the benchmarks were conducted on a 64bit Intel Pentium VI 
Quad CPU 2.40 GHZ machine with 4 MB cache and 3 GB main memory.

For the solving of bivariate systems we compare
the command {\tt Trian\-gularize} of the {\RC}
library to the command {\tt BivariateModularTri\-angularize}
of the module {\tt FastArithmeticTools}.
Indeed both commands have the same specification
for such input systems.
Note that {\tt Trian\-gularize} is a high-level generic code 
which applies to any type of input system
and which does
not rely on fast polynomial arithmetic or modular methods.
On the contrary, {\tt BivariateModularTri\-angularize}
is specialized to bivariate systems 
(see Section~\ref{sec:ResultantAndGCSolvingTwoEquations}
and Corollary~\ref{prop:OurAlgorithm2})
is mainly implemented in C
and is supported by the {\Modpn} library.
{\tt BivariateModularTri\-angularize} is an instance
of a more general fast algorithm called
{\tt FastTrian\-gularize}; we use this second name
in our figures.

Since a triangular decomposition can be regarded 
as a ``factored'' lexicographic Gr\"obner basis
we also benchmark the computation of such bases
in {\Maple} and {\Magma}.

\begin{figure}[h]
  \begin{center}
    \includegraphics[scale=0.5]{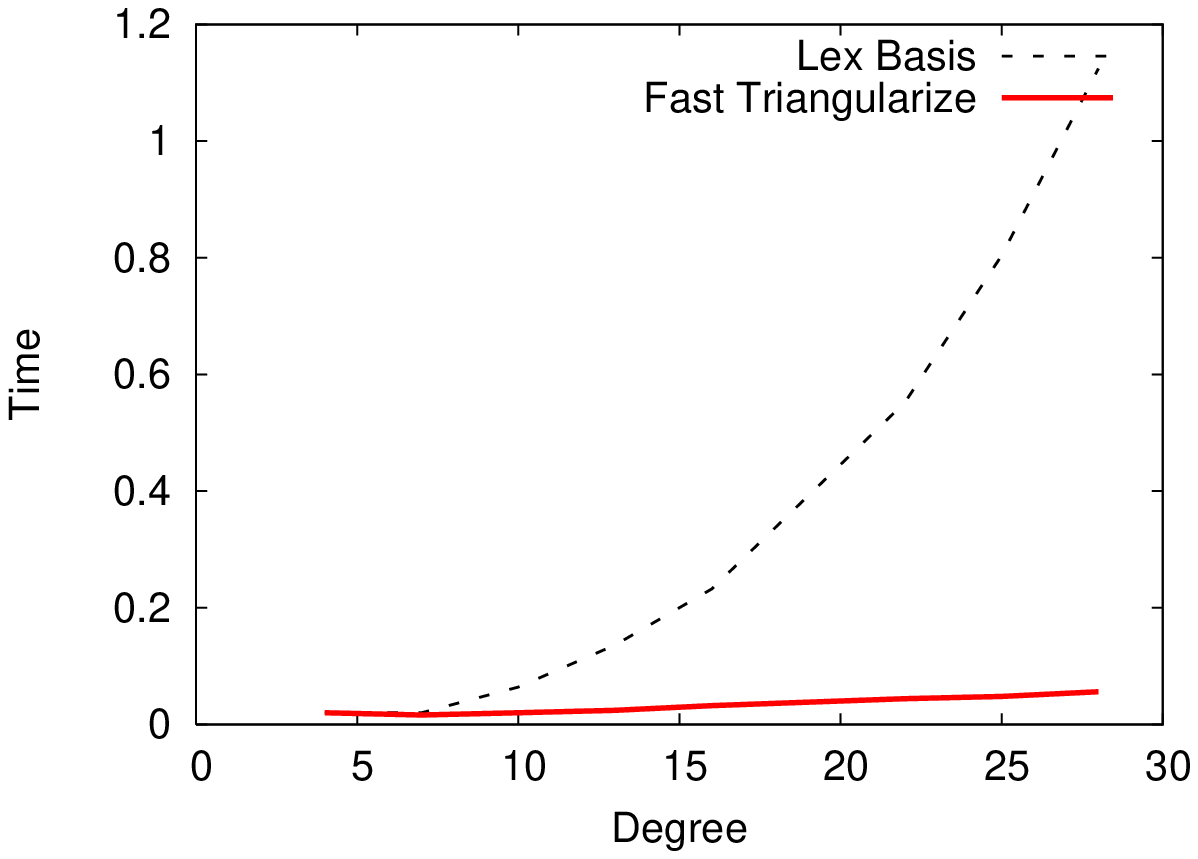}
   \caption{{\scriptsize Generic dense bivariate systems.}}
    \label{DenseSolve2vsMapleSub}
  \end{center}
\end{figure}

\begin{figure}[h]
  \begin{center}
    \includegraphics[scale=0.5]{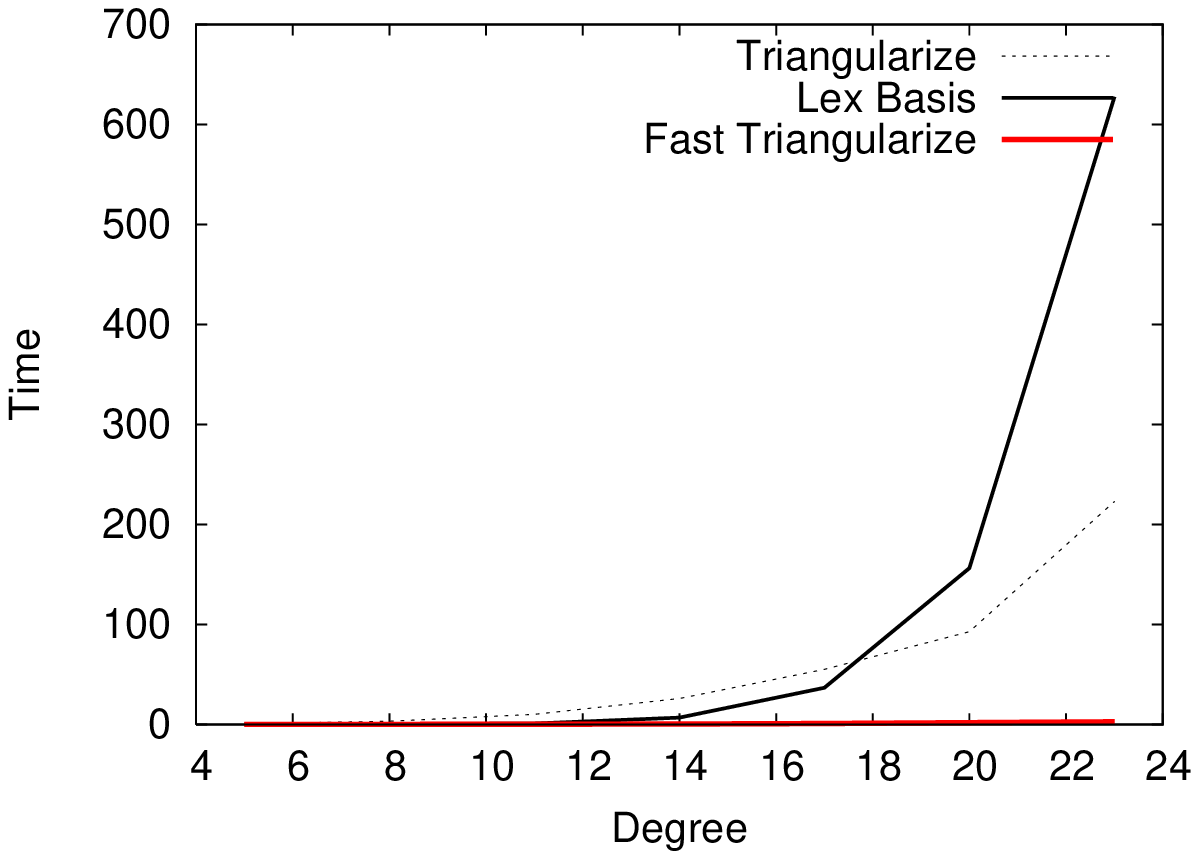}
   \caption{{\scriptsize Highly non-equiprojectable bivariate systems.}}
    \label{SplitSolve2vsMaple}
  \end{center}
\end{figure}

\begin{figure}[h]
  \begin{center}
    \includegraphics[scale=0.5]{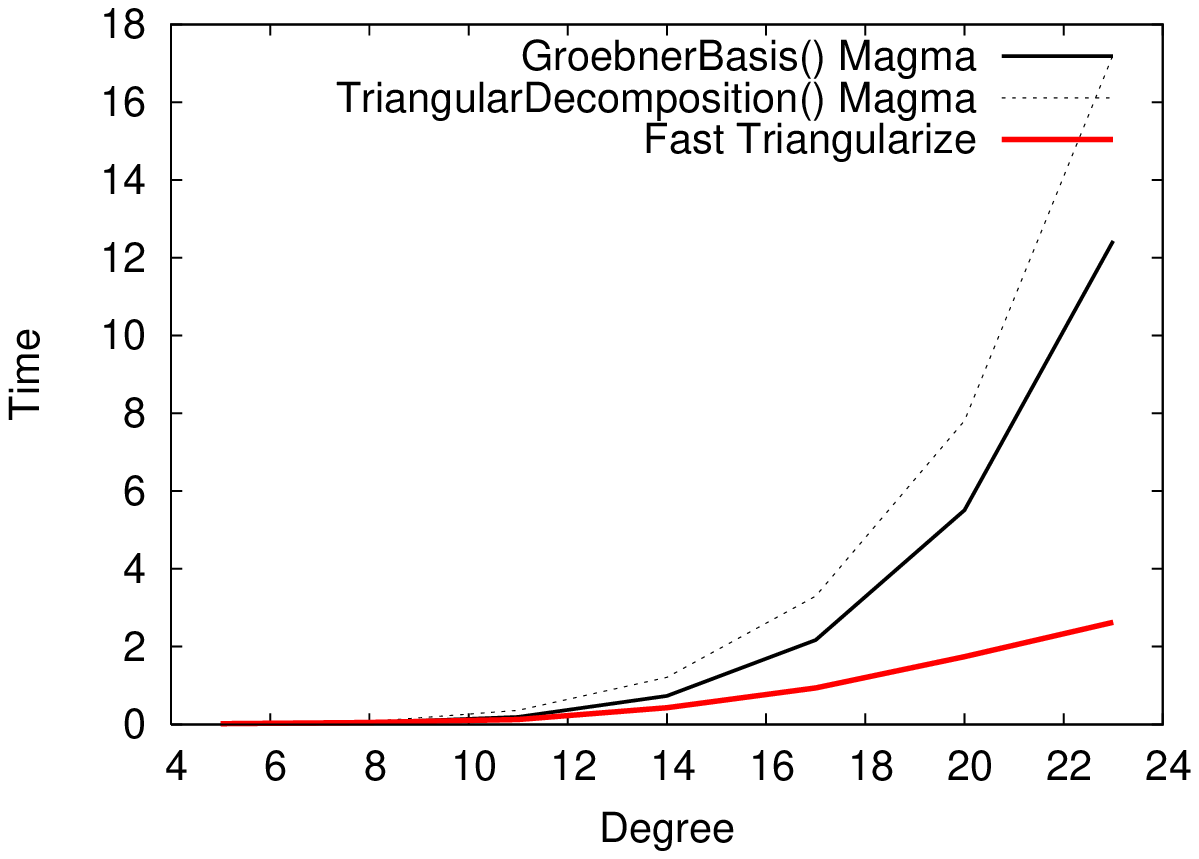}
   \caption{{\scriptsize Highly non-equiprojectable bivariate systems.}}
    \label{SplitSolve2vsMagma}
  \end{center}
\end{figure}

Figure~\ref{DenseSolve2vsMapleSub} compares 
{\tt FastTrian\-gularize} and (lexicographic)
{\tt Groebner:-Basis} in {\Maple} on generic
dense input systems.
On the largest input example the former solver
is about 20 times faster than the latter.
Figure~\ref{SplitSolve2vsMaple} compares
{\tt FastTrian\-gularize} and (lexicographic)
{\tt Groebner:-Basis} on highly non-equiprojectable
dense input systems; for these systems the number
of equiprojectable components is about half the degree
of the variety.
At the total degree 23 our solver is 
approximately 100 times faster than {\tt Groebner:-Basis}.
Figure~\ref{SplitSolve2vsMagma}
compares
{\tt FastTrian\-gularize},
{\tt GroebnerBasis} in {\Magma}
and
{\tt TriangularDecomposition}  in {\Magma}
on the same set of highly non-equiprojectable
dense input systems.
Once again our solver outperforms its competitors.


\begin{figure}[h]
  \begin{center}
    \includegraphics[scale=0.5]{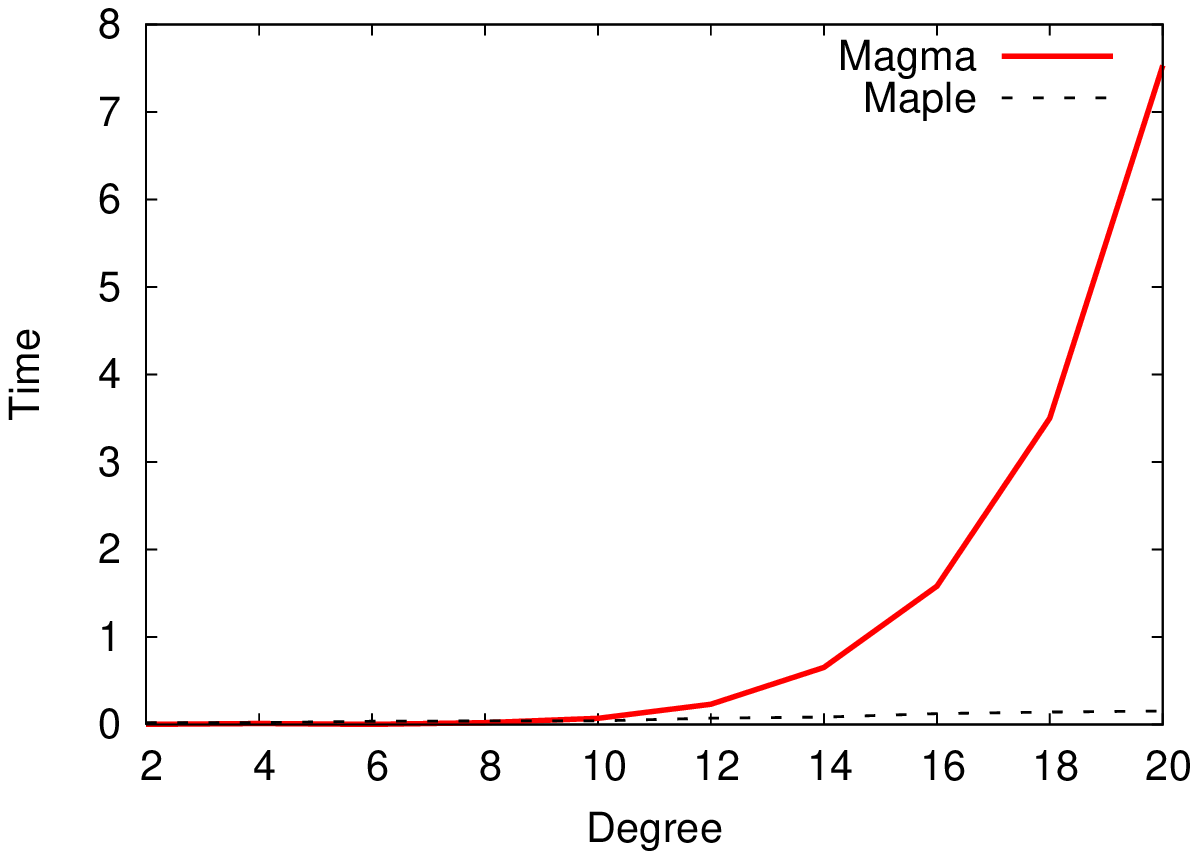}
   \caption{{\scriptsize {\em Generic dense trivariate systems.}}}
    \label{DenseResGcd2VarDFTvsMagmaB}
  \end{center}
\end{figure}

For the solving of systems with two equations,
we compare {\tt FastTriangularize} 
(implementing in this case 
 the algorithm described in 
Section~\ref{sec:ResultantAndGCSolvingTwoEquations})
with {\tt GroebnerBasis} in {\Magma}.
On Figure~\ref{DenseResGcd2VarDFTvsMagmaB} 
these two solvers are simply referred as {\Magma}
and {\Maple}.
For this benchmark the input systems are generic
dense trivariate systems.


Figures~\ref{Tab:InverseNonSplit2},
\ref{Tab:InverseNonSplit3}
and
\ref{Tab:InverseSplit3}
compare our fast regularity test algorithm
(Algorithm~\ref{Algo:IsInvertible})
with the {\RegularChains} library {\tt Regula\-rize}
and its {\Magma} counterpart.
More precisely, in  {\Magma}, we first saturate the ideal generated by 
the input zerodimensional regular chain $T$ 
with the input polynomial $P$ using  
the {\tt Saturation} command. 
Then the {\tt Triangu\-lar\-De\-composition} 
command decomposes the output produced by the first step.  
The total degree of the input $i$-th polynomial in $T$ is $d_i$. 
For Figure~\ref{Tab:InverseNonSplit2} and Figure~\ref{Tab:InverseNonSplit3}
the input $T$ and $P$ are random such that the intermediate
computations do not split.
In this ``non-splitting'' cases, 
our fast {\tt Regularize} algorithm is 
significantly faster than the other commands. 

For Figure~\ref{Tab:InverseSplit3}
the input $T$ and $P$ are built such that many intermediate
computations need to split.
In this case, our fast {\tt Regularize} algorithm
is slightly slower than its {\Magma} counterpart, 
but still much faster than 
the ``generic'' (non-modular and non-supported by {\Modpn})
{\tt Regularize} command of the {\RegularChains} library.
The slow down w.r.t. the {\Magma} code is due
to the (large) overheads of the C - {\Maple} interface,
see \cite{LMRS08} for details.

\begin{figure}[htbp]
    \begin{center}
    {\scriptsize
        \begin{tabular}{| r | r | r | r | r |}  \hline
          $d_1$ & $d_2$ & Regularize & Fast Regularize &   Magma \\ \hline
             2  & 2     &    0.052   & 0.016           &   0.000 \\ \hline
             4  & 6     &    0.236   & 0.016           &   0.010 \\ \hline
             6  & 10    &    0.760   & 0.016           &   0.010 \\ \hline
             8  & 14    &    1.968   & 0.020           &   0.050 \\ \hline
             10 & 18    &    4.420   & 0.052           &   0.090 \\ \hline
             12 & 22    &    8.784   & 0.072           &   0.220 \\ \hline
             14 & 26    &   15.989   & 0.144           &   0.500 \\ \hline
             16 & 30    &   27.497   & 0.208           &   0.990 \\ \hline
             18 & 34    &   44.594   & 0.368           &   1.890 \\ \hline
             20 & 38    &   69.876   & 0.776           &   3.660 \\ \hline
             22 & 42    &  107.154   & 0.656           &   6.600 \\ \hline
             24 & 46    &  156.373   & 1.036           &  10.460 \\ \hline
             26 & 50    &  220.653   & 2.172           &  17.110 \\ \hline
             28 & 54    &  309.271   & 1.640           &  25.900 \\ \hline
             30 & 58    &  434.343   & 2.008           &  42.600 \\ \hline
             32 & 62    &  574.923   & 4.156           &  57.000 \\ \hline
             34 & 66    &  746.818   & 6.456           & 104.780 \\ \hline
        \end{tabular}
        \caption{\scriptsize 2-variable random dense case.}
        \label{Tab:InverseNonSplit2}
    }
    \end{center}
\end{figure}

\begin{figure}[htbp]
    \begin{center}
    {\scriptsize
        \begin{tabular}{| r | r | r | r | r | r |}  \hline
        $d_1$ & $d_2$ & $d_3$ & Regularize & Fast Regularize &   Magma \\ \hline
            2 &  2    &  3    &    0.240   &    0.008        &   0.000 \\ \hline
            3 &  4    &  6    &    1.196   &    0.020        &   0.020 \\ \hline
            4 &  6    &  9    &    4.424   &    0.032        &   0.030 \\ \hline
            5 &  8    & 12    &   12.956   &    0.148        &   0.200 \\ \hline
            6 & 10    & 15    &   33.614   &    0.360        &   0.710 \\ \hline
            7 & 12    & 18    &   82.393   &    1.108        &   2.920 \\ \hline
            8 & 14    & 21    &  168.910   &    2.204        &   8.250 \\ \hline
            9 & 16    & 24    &  332.036   &   14.764        &  23.160 \\ \hline
           10 & 18    & 27    &  $>$1000   &   21.853        &  61.560 \\ \hline
           11 & 20    & 30    &  $>$1000   &   57.203        & 132.240 \\ \hline
           12 & 22    & 33    &  $>$1000   &  102.830        & 284.420 \\ \hline
        \end{tabular}
        \caption{\scriptsize 3-variable random dense  case.}
        \label{Tab:InverseNonSplit3}
    }
    \end{center}
\end{figure}

\begin{figure}[ht]
    \begin{center}
    {\scriptsize
        \begin{tabular}{| r | r | r | r | r | r | r |}  \hline
        $d_1$  & $d_2$ & $d_3$ & Regularize &  Fast Regularize  &  {\tt V2.15-4}  &  {\tt V2.14-8}  \\ \hline
             2 &     2 &    3  & 0.184      &          0.028    &    0.000     &  0.000       \\ \hline
             3 &     4 &    6  & 0.972      &          0.060    &    0.000     &  0.010       \\ \hline
             4 &     6 &    9  & 3.212      &          0.092    &  $>$1000     &  0.030       \\ \hline
             5 &     8 &    12 & 8.228      &          0.208    &  $>$1000     &  0.150       \\ \hline
             6 &    10 &    15 & 21.461     &          0.888    &  807.850     &  0.370       \\ \hline
             7 &    12 &    18 & 51.751     &          3.836    &  $>$1000     &  1.790       \\ \hline
             8 &    14 &    21 & 106.722    &          9.604    &  $>$1000     &  2.890       \\ \hline
             9 &    16 &    24 & 207.752    &         39.590    &  $>$1000     & 10.950       \\ \hline
            10 &    18 &    27 & 388.356    &         72.548    &  $>$1000     & 19.180       \\ \hline
            11 &    20 &    30 & 703.123    &        138.924    &  $>$1000     & 56.850       \\ \hline
            12 &    22 &    33 & $>$1000    &        295.374    &  $>$1000     & 76.340       \\ \hline
             \end{tabular}
             \caption{\scriptsize 3-variable dense case with many splittings.}
             \label{Tab:InverseSplit3}
    }
    \end{center}
\end{figure}

\section{Conclusion}
\label{sec:conclusion}
The concept of a regular GCD extends the usual notion of polynomial GCD 
from polynomial rings over fields
to polynomial rings modulo saturated ideals of regular chains.
Regular GCDs play a central role in triangular decomposition methods.
Traditionally, regular GCDs are
computed in a top-down manner, by adapting standard PRS techniques
(Euclidean Algorithm, subresultant algorithms, \ldots).

In this paper, we have examined the properties of regular GCDs of
two polynomials w.r.t a regular chain. The theoretical results
presented in Section~\ref{sec:RegularGcds} 
show that one can proceed in a bottom-up manner.
This has three benefits described in 
Section~\ref{sec:ImplementationAndComplexity}.
First, this algorithm is well-suited to employ modular methods and 
fast polynomial arithmetic.
Secondly, we avoid the repetition of (potentially expensive)
intermediate computations.
Lastly, we avoid, as much as possible, computing
modulo regular chains and use polynomial computations
over the base field instead, while controlling expression swell.
The experimental results reported in Section~\ref{sec:experimentation}
illustrate the high efficiency of our algorithms.

\section{Acknowledgement}
The authors would like to thank our friend Changbo Chen, who pointed out
that Lemma $3$ in an earlier version of this paper is incorrect.
\nocite{LMPX08}

\bibliographystyle{plain}

\begin{thebibliography}{10}
\bibitem{AM69}
M.F. Atiyah and L.~G. Macdonald.
\newblock {\em Introduction to Commutative Algebra}.
\newblock Addison-Wesley, 1969.

\bibitem{BLM06}
F.~Boulier, F.~Lemaire, and M.~{Moreno Maza}.
\newblock {Well known theorems on triangular systems and the {D5} principle}.
\newblock In {\em Proc. of {\em Transgressive Computing 2006}}, Granada, Spain,
  2006.

\bibitem{CantorKaltofen1991}
D.G. Cantor and E.~Kaltofen.
\newblock On fast multiplication of polynomials over arbitrary algebras.
\newblock {\em Acta Informatica}, 28:693--701, 1991.

\bibitem{CLGMP07}
C.~Chen, O.~Golubitsky, F.~Lemaire, M.~{Moreno Maza}, and W.~Pan.
\newblock {\em Comprehensive Triangular Decomposition}, volume 4770 of {\em
  LNCS}, pages 73--101.
\newblock Springer Verlag, 2007.

\bibitem{Coll71}
G.E. Collins.
\newblock The calculation of multivariate polynomial resultants.
\newblock {\em Journal of the {ACM}}, 18(4):515--532, 1971.

\bibitem{DMSX06}
X.~{Dahan}, M.~{{Moreno Maza}}, {\'E}.~{Schost}, and Y.~{Xie}.
\newblock On the complexity of the {{D5}} principle.
\newblock In {\em Proc. of {\em Transgressive Computing 2006}}, Granada, Spain,
  2006.

\bibitem{D5}
J.~{Della Dora}, C.~Dicrescenzo, and D.~Duval.
\newblock About a new method for computing in algebraic number fields.
\newblock In {\em Proc. EUROCAL 85 Vol. 2}, Springer-Verlag, 1985.

\bibitem{Duc97}
L.~Ducos.
\newblock {\em Effectivit\'e en th\'eorie de Galois. Sous-r\'esultants}.
\newblock PhD thesis, Universit{\'e} de Poitiers, 1997.

\bibitem{DD87}
D.~Duval.
\newblock {\em Questions Relatives au Calcul Formel avec des Nombres
  Alg{\'e}briques}.
\newblock Universit{\'e} de {Grenoble}, 1987.
\newblock Th{\`e}se d'{\'E}tat.

\bibitem{ElKahoui2003}
{M'hammed} {El Kahoui}.
\newblock An elementary approach to subresultants theory.
\newblock {\em J. Symb. Comp.}, 35:281--292, 2003.

\bibitem{GaGe99}
{J. von zur} Gathen and J.~Gerhard.
\newblock {\em Modern Computer Algebra}.
\newblock Cambridge University Press, 1999.

\bibitem{Gom92}
T.~G{\'o}me{z D{\'{\i}}az}.
\newblock {\em Quelques applications de l'{\'e}valuation dynamique}.
\newblock PhD thesis, Universit{\'e} de Limoges, 1994.

\bibitem{Hoeven04}
{J. van der} Hoeven.
\newblock The {T}runcated {F}ourier {T}ransform and applications.
\newblock In {\em ISSAC'04}, pages 290--296. ACM, 2004.

\bibitem{Kalk93}
M.~Kalkbrener.
\newblock A generalized euclidean algorithm for computing triangular
  representations of algebraic varieties.
\newblock {\em J. Symb. Comp.}, 15:143--167, 1993.

\bibitem{Laz91a}
D.~Lazard.
\newblock A new method for solving algebraic systems of positive dimension.
\newblock {\em Discr. App. Math}, 33:147--160, 1991.

\bibitem{Laz92}
D.~Lazard.
\newblock Solving zero-dimensional algebraic systems.
\newblock {\em J. Symb. Comp.}, 15:117--132, 1992.

\bibitem{LMPX08}
F.~Lemaire, M.~{Moreno Maza}, W.~Pan, and Y.~Xie.
\newblock When does $(T)$ equal ${\rm Sat}(T)$?
\newblock In {\em Proc. ISSAC'20008}, pages 207--214. ACM Press, 2008.

\bibitem{LeMoXi05}
F.~Lemaire, M.~{Moreno Maza}, and Y.~{Xie}.
\newblock The {{{\tt RegularChains}}} library.
\newblock In {Ilias S. Kotsireas}, editor, {\em {\em Maple Conference 2005}},
  pages 355--368, 2005.

\bibitem{LMRS08}
X.~Li, M.~{Moreno Maza}, R.~Rasheed, and {\'E}.~Schost.
\newblock The modpn library: Bringing fast polynomial arithmetic into maple.
\newblock In {\em MICA'08}, 2008.

\bibitem{LMS07}
X.~Li, M.~{Moreno Maza}, and {\'E}.~Schost.
\newblock Fast arithmetic for triangular sets: From theory to practice.
\newblock In {\em ISSAC'07}, pages 269--276. ACM Press, 2007.

\bibitem{Mis93}
B.~Mishra.
\newblock {\em Algorithmic Algebra}.
\newblock Springer, New York, 1993.

\bibitem{MMM99}
M.~{{Moreno Maza}}.
\newblock On triangular decompositions of algebraic varieties.
\newblock Technical Report TR {4/99}, NAG Ltd, Oxford, UK.
\newblock Presented at the MEGA-2000 Conference, Bath, England.

\bibitem{MMMRR}
M.~{{Moreno Maza}} and R.~Rioboo.
\newblock Polynomial gcd computations over towers of algebraic extensions.
\newblock In {\em Proc. AAECC-11}, pages 365--382. Springer, 1995.

\bibitem{Pan94}
{V.~Y.} Pan.
\newblock Simple multivariate polynomial multiplication.
\newblock {\em J. Symb. Comp.}, 18(3):183--186, 1994.

\bibitem{Yap1993}
{C.~K.} Yap.
\newblock {\em Fundamental Problems in Algorithmic Algebra}.
\newblock Princeton University Press, 1993.

\end{thebibliography}

\end{document}